\documentclass[pre,noshowpacs,noshowkeys,twocolumn]{revtex4}
\usepackage{graphicx}
\usepackage{bm}
\usepackage{textcomp}
\usepackage{color}
\usepackage{amsfonts}

\begin{document}

\title{Macroscopic Floquet topological crystalline steel pump}
\author{Anna M. E. B. Rossi$^{a}$, Jonas Bugase$^{a}$ and Thomas M. Fischer$^{a}$}\email{thomas.fischer@uni-bayreuth.de}
\affiliation{$^{a}$ Experimental Physics, Institute of Physics and Mathematics, Universit\"at Bayreuth, 95440 Bayreuth (Germany).
}
\date{\today}

\begin{abstract}
	The transport of a steel sphere on top of two dimensional periodic 
	magnetic patterns is studied experimentally. 
	Transport of the sphere is achieved by moving an external permanent magnet on a closed loop around the two dimensional crystal.
	The transport is topological i.e. the steel sphere is transported by a primitive unit vector of the lattice  when the external magnet loop winds around specific directions. We experimentally determine the set of directions the loops must enclose for nontrivial transport of the steel sphere into various directions. \end{abstract}

\maketitle

\section{Introduction}\label{Introduction}
Topological non trivial matter is a class of material, where the response of the material to external perturbations only depends on global properties not on local properties of the material. Such properties are called topological invariants and they change in a discrete way, i.e. a continuous change of the perturbation results in a discrete response of the material. Topological properties of matter play a fundamental role in electronic transport behavior of quantum solid state matter \cite{Hasan,TI}, in  mesoscopic systems \cite{Rechtsman,Mao, Murugan,Loehr,delasHeras,Loehr2} and in macroscopic matter \cite{Kane,Paulose,Nash,Huber}. One important class of topological material are Floquet topological systems, where the material is subject to a time periodic external perturbation that causes the pumping of excitations or quasi particles through the material. The topological pump effect \cite{Mele} is usually protected by certain symmetries of the problem. One of such symmetries are the point group symmetries of the lattice.

The current work presents three macroscopic examples of topological magnetic crystals, with magnetic point symmetry protected Floquet transport properties of paramagnetic (soft magnetic) spheres placed above the crystal. We experimentally determine regions of orientation we have to wind around an external magnetic field to pump the spheres into certain directions.    

\section{steel pump setup}

\begin{figure*}
	\includegraphics[width=2\columnwidth]{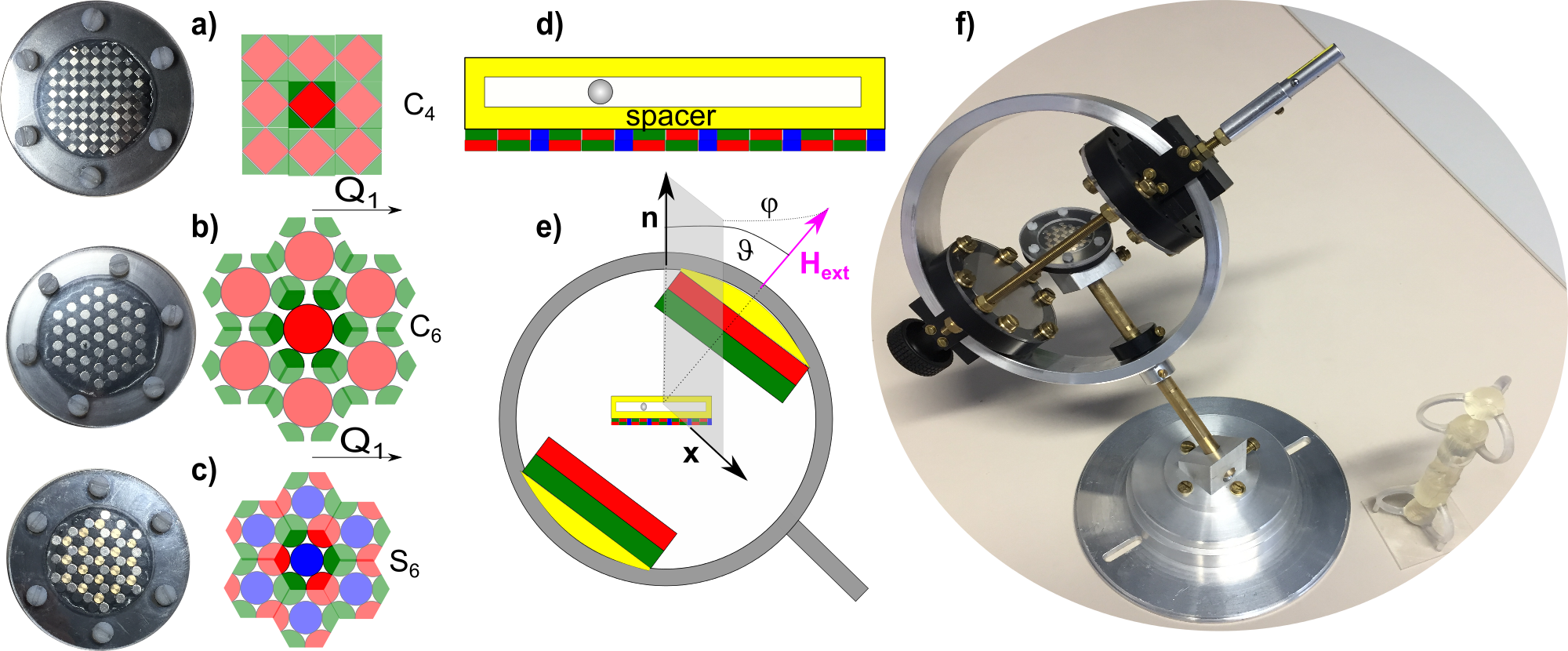}
	\caption{a) Top view of the magnetic pattern of symmetry $C_4$, sample left and scheme right b) symmetry $C_6$ and c) symmetry $S_6$.
		Silver areas in the sample (red areas in the scheme) are magnetized up and black (green) areas are magnetized down respectively. Blue areas are non magnetic brass cylinders inserted for mechanical stability. One unit cell is emphasized in full colors. The vector $\mathbf{Q}_1$ is one of the primitive reciprocal lattice vectors. d) Sideview of the pattern and the compartment holding the steel sphere. e) Goniometer and external magnets surrounding the sample. f) A photo of the setup.}
	\label{figurescheme}
\end{figure*}

Three topological two dimensional magnetic crystals are built from an arrangement of NbB-magnets. The first crystal consists of magnetic cubes of side length $d_1=2mm$ and remanence $\mu_0 M_1=1.35 T$ arranged in a four fold symmetric $C_4$ checkerboard square lattice of alternating up and down magnetized cubes (Fig. 1a). The second, a hexagonal lattice (Fig. 1b) consists of cylindrical magnets of diameters $d_2=3 mm$ and $d_3=2 mm$, height $h=2mm$ and remanence $\mu_0 M_2= 1.19 T$  and $\mu_0 M_3= 1.35 T$ (respectively). The larger size $d_2=3mm$ magnets are magnetized upwards and they are surrounded by six smaller size $d_3=2mm$ magnets that are magnetized downwards and that touch the larger magnet. The primitive unit cell of the lattice is a six fold symmetric $C_6$ hexagon with corners centered within the smaller magnets. Each unit cell thus contains one large magnet and two smaller magnets. The third lattice (Fig. 1c) is built from a hexagonally closed packed arrangement of $d_2=3mm$ diameter cylinders. The central cylinder (blue) is non magnetic brass, and the surrounding cylinders are NbB magnets of alternating up and downward magnetization creating an improper six fold $S_6$ symmetry. 
All two dimensional lattices have two primitive lattice vectors of the same length $a_1=a_2=a=2.82 mm$ (square lattice), $a=2.5 mm$ ($C_6$ lattice), and  $a=5.2 mm$ ($S_6$ lattice) and are metastable (the ground state configuration of the magnet ensemble is a magnetic rod of magnets aligned along one axis) in zero external magnetic field. We fix the arrangement with an epoxy resin placed in the voids and the two dimensional surroundings of the pattern.
The pattern then is stable also in the presence of an external field. The crystals are put on a support and covered with a transparent PMMA spacer of thickness $z=1-1.5mm$ (Fig. 1d). We place a steel sphere of diameter $2r=1 mm$ on top of the spacer and create a closed but transparent compartment around the steel sphere. The topological magnetic crystal with the steel sphere on top is placed in the center of a goniometer set up at an angle of 45 degrees to ensure that relevant motion is not effected by the restrictions of motion of the goniometer (Fig. 1e and f) caused by the support. The goniometer holds two NbB-magnets of diameter $d_{ext}=60mm$, thickness  $t_{ext}=10 mm$ and remanence $\mu_0 M_{ext}=1.28 T$ aligned parallel to each other at a distance $2R= 120 mm$ and creating an external magnetic field $H_{ext}=3600 A/m$ penetrating the two dimensional crystal and the steel sphere. The magnetic field gradients ${\boldmath \nabla H_{ext}} \approx M_{ext}t_{ext}d_{ext}^2/R^4$ of the external field at the position of the steel sphere is at least two orders of magnitude smaller than the field gradients of the magnetic field of the crystal ${\boldmath \nabla H_{int}} \approx M/a$. The two external magnets can be oriented to produce an arbitrary direction of the external magnetic field with respect to the crystal. A laser pointer pointing along ${\mathbf H}_{ext}$ is mounted on the goniometer creating a stereographic projection of the instantaneous external magnetic field direction on a recording plane. 

\begin{figure*}
	\includegraphics[width=2\columnwidth]{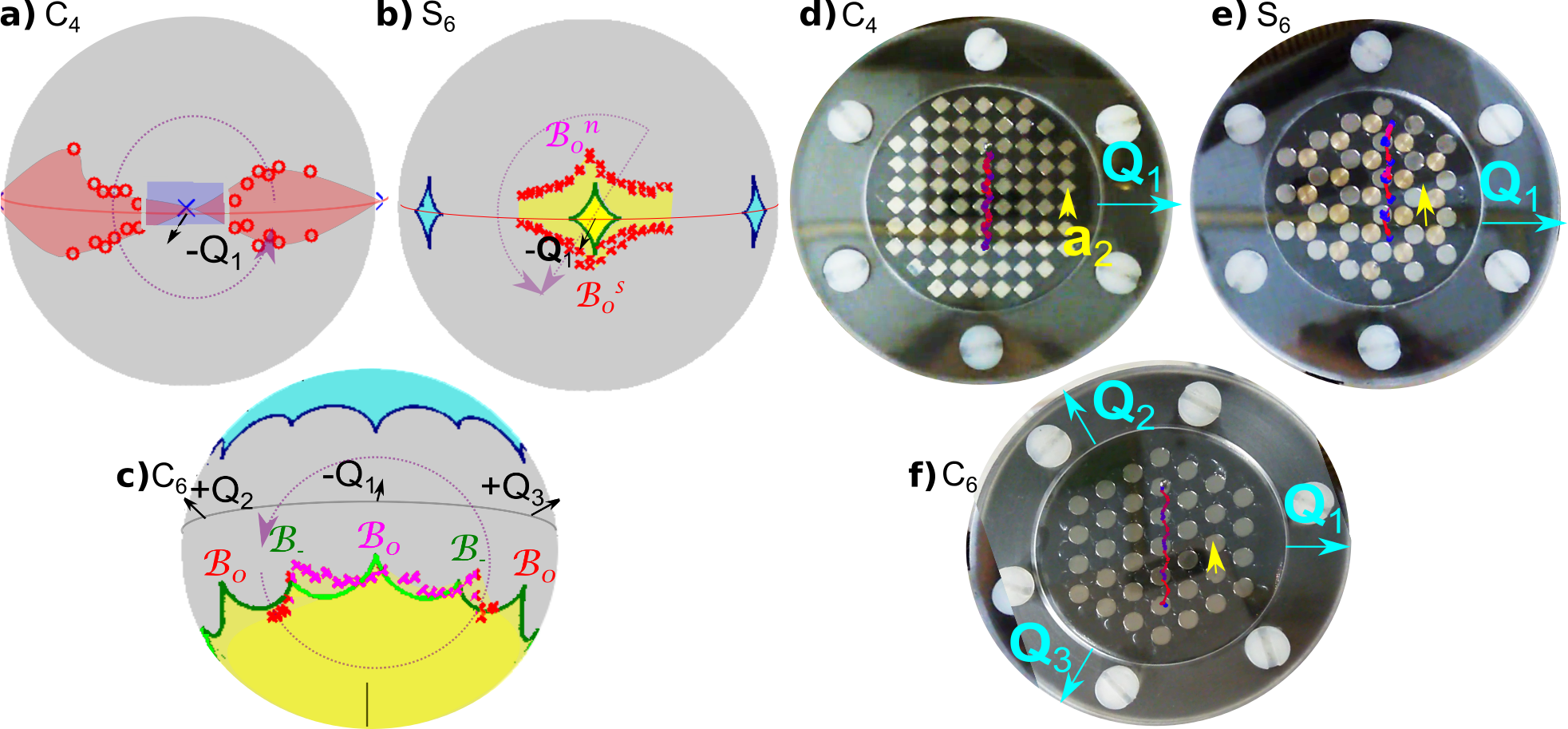}
	\caption{Control spaces of a) the $C_4$-symmetric, b) the $S_6$-symmetric, and c) the $C_6$-symmetric lattice. The $-\mathbf{Q}_1$-direction of control space is opposite to the reciprocal lattice vector direction $\mathbf{Q}_1$ in d). a) theoretical fence points are shown in blue, gates as red lines. The fence points that one must wind around to achieve non-trivial transport enlarges to the fence area (blue) in the experiment. Gates are shown as red circles and show a hystereses (red area) when winding around the fence area in different directions. We depict a purple loop encircling the blue area as an  example loop inducing non-trivial colloidal transport into the $\mathbf{n\times Q}_1$-direction. b) theoretical fences for paramagnets are shown in green and for diamagnets in blue. The experimentally determined fence for the steel sphere lies further outside with two separate regions of instability when leaving the yellow area toward the north or south. We depict a palindrome modulation loop in purple that cycles through the common area back and forth in control space but causes an open trajectory with ratchet jumps of the steel sphere above the lattice. For the part of the loop moving in the mathematical positive sense the winding number around ${\cal B}_0^n$ is non-zero and causes nontrivial transport, while for the same loop traveling in the mathematical negative sense the winding number around ${\cal B}_0^s$ is zero and causes trivial motion c) theoretical fences as green and blue lines with the experimental fence shown as pink and red crosses. The purple example loop causes adiabatic transport in the $(\mathbf{Q}_2-\mathbf{Q}_3)$-direction. d-f) Trajectories (blue and red) of the steel sphere subject to the purple loops shown in a-c. Red segments correspond to faster motion than blue segments. The adiabatic motion d) and f) smoothly changes from fast to slower, while the ratchet motion e) discontinuously  changes from slow adiabatic to fast ratchet jumps. The yellow arrow corresponds to the primitive unit vector pointing into the transport direction.  Movies of the motion are provided in \cite{EPAPS}.}
	\label{figurescheme}
\end{figure*}

\section{Topologically nontrivial transport loops}

We reorient the external magnets by moving along a closed reorientation loop that starts and ends at the same initial orientation. The steel sphere responds to the reorientation loop with a motion that starts at one position of the lattice and ends at a final position. A topological trivial motion of the steel sphere is a motion where the steel sphere responds to a closed reorientation loop with a closed loop on the lattice. Not every closed reorientation loop causes a trivial response of the steel sphere. There are  topologically non trivial trajectories, where the steel sphere trajectory ends at a position differing from the initial position by one vector of the lattice.

We choose a collection of different non self intersecting reorientation loops and measured the corresponding displacement of the steel sphere. Each non self intersecting reorientation loop cuts the sphere of orientations that we call the control space into two areas. One of the areas is circulated by the loop in the positive sense  the other in the negative sense. We define the intersection of all positive areas of loops causing the same net transport of the steel sphere as the positive common area.  Similarly we can define the negative common area of the same transport directions.    

In Fig. 2a we show the common area (blue) determined in this way for the transport into the $\mathbf{n\times Q}_1$-direction for the fourfold symmetric pattern. The common area is a rectangle centered around the primitive reciprocal vector $-\mathbf{Q}_1$ of the lattice. Whenever we wind the modulation loop around the common area in a way that does not touch the area the result is the same non-trivial transport. Entering the common area leads to a statistical trivial or non-trivial response transport direction of the steel sphere. The sphere passes from the up-magnetized region toward the down-magnetized regions or vice versa when the loop crosses the gates (red circles). In the experiments we observe a hysteresis, i.e. the gate in control space is positioned at the red circles of the southern hemisphere when the external field moves from north toward the south and the northern hemisphere for the opposite direction. The region of the hysteresis is shown as the red area in control space. Note that similar common areas repeat every $2\pi/4$ along the equator because of the $C_4$-symmetry. In previous work \cite{Loehr2} we have computed the theoretical position of the common area as well as the position of the gates. Theoretically the common area is just one point, the $-\mathbf{Q}_1$-direction, and the gate is a (red) line on the equator showing no hysteresis.

In Fig. 2b we show the control space of the $S_6$-symmetric patterns. Non-trivial transport into the $\mathbf{n\times Q}_1$-direction occurs if we wind the loop around the yellow common area. We call the borders of the yellow area the fence. A new feature of the $S_6$-symmetric pattern is that the transport is still predictable if we enter and exit the yellow area with a loop through fence segments marked in red. A loop exiting the common area in the north (south) has the same result as a common area avoiding loop with the same winding number around the ${\cal B}_0^n$ (${\cal B}_0^s$) point. Although the transport direction of those common area passing loops are the same as the common area avoiding loops, their character is that of a ratchet. A jump of the steel sphere from one point on the lattice to a different point occurs when we exit the common area. Loops avoiding the common area cause a smooth quasi adiabatic transport of the steel sphere. Entering or exiting the common area in regions where there is no red fence segments yields statistical results for the steel transport direction. Two further common areas exist at the location turned by $\pm 2\pi/3$ along the equator. The theoretical prediction is in topological agreement with the experiments, however, the common area enclosed between the green fence segments is smaller than the experimentally measured area and the northern and southern fence segments form a closed line around it with no statistical segments of the common area border. The theoretical common areas repeat when we turn the control space by $2\pi/3$ around the normal vector. The cyan common areas correspond to theoretical areas supposed to cause non-trivial transport of diamagnetic (superconducting) spheres. 

In Fig. 2c we show the control space of the $C_6$-symmetric pattern. The transport is adiabatic if we encircle an even number of ${\cal B}_-$ points, i.e. when entering and exiting segments that have the same pink or red (bright or dark green) color, and of ratchet character otherwise. Non trivial transport into the $\sigma( \mathbf{ Q}_i-\mathbf{Q}_j)$-direction ($\sigma=\pm 1$, $i,j=1,2,3$) occurs when the modulation loop enters the yellow area via a neighbor segment of the reciprocal unit vector $\sigma\mathbf{ Q}_i$ and exits via a nearest or next nearest neighbor segment of the reciprocal lattice vector $\sigma\mathbf{Q}_j$.  The sign $\sigma$ of the nearest or next nearest exit reciprocal vector $\sigma\mathbf{Q}_j$ must be the same as that of nearest reciprocal vector $\sigma\mathbf{ Q}_i$ of the entry. The experimental position of the fence (pink and red) has been determined from the irreversible jumps of the steel sphere when the external field exits the yellow area near a reciprocal lattice vector having opposite sign to the reciprocal lattice vector of the entry. The match between experiment and theory here is almost perfect.  


In Fig. 2d we show the trajectory of the steel sphere subject to a purple loop in Fig. 2a encircling the reciprocal vector $-\mathbf{Q}_1$ in the mathematical positive sense. A movie of the motion can be found in \cite{EPAPS}. In Fig. 2e we show the trajectory of a palindrome modulation loop for the $S_6$-symmetric lattice shown in Fig. 2b. The loop crosses the common area by entering the left southern fence segment and exiting at the northern right segment and returning to the initial orientation left of the common area. Immediately afterwards the palindrome loop retraces the path in the opposite direction. The trajectory of the steel sphere is of the ratchet type and does not close because of irreversible jumps that happen when the modulation loop leaves the common area at different fence segments during the forward and backward period. Finally in Fig. 2f we depict the adiabatic trajectory of the steel sphere above a $C_6$-symmetric lattice for a loop passing through the yellow area via the red fence segments in Fig 2c.

\section{Discussion and Conclusion}
From the measurements we see that the experiments are in topological agreement with the theory \cite{Loehr,delasHeras,Loehr2}. The most striking difference between experiment and theory is the existence of a hysteresis visible in the $C_4$- and $S_6$- symmetric patterns. The theory assumes that particles will jump from an instable position toward the new equilibrium position at the fence, where the curvature of the potential minimum vanishes. We explain the shift of the experimental fence with respect to the theoretical predictions as well as the hysteresis by solid friction that lets the steel particle move only when the magnetic potential exceeds a certain slope. Slopes for a forward and backward jump will have opposite sign explaining the splitting of the closed theoretical fence in the $S_6$-pattern into two separate forward and backward loop fences. 

Let us note that the topological protected transport theory has been developed for colloidal particles not for steel spheres. The scale invariance of the theory demonstrates the robustness of the topological concept. Presumably it is also possible to down scale the experiment from the colloidal toward molecular scales, which would provide a transport mechanism for molecular magnets above magnetic nano structures.

\section{acknowledgments}
J. B  acknowledges financial support by a Ghana MOE - DAAD joined fellowship. We appreciate scientific support by Johannes Loehr and by Daniel de las Heras.

\end{document}